# LIMIT THEOREMS FOR THE DISORDERED QUANTUM WALK

**CLEMENT AMPADU**

31 Carrolton Road
Boston, Massachusetts, 02132
USA
e-mail: drampadu@hotmail.com

**Abstract**

We study the disordered quantum walk in one dimension, and obtain the weak limit theorem.

I. Introduction

Limit theorems for quantum walks has been well studied by many authors [1-21], for example. In this paper we study the disordered quantum walk as defined by N.Konno [22]. We should remark that the unitary transformation governing our walk is an example of a *disordered quantum walk of type II* as defined Section II. The matrix was studied by Mackay et.al [23] in their analysis of quantum walk in higher spatial dimensions, comparing classical and quantum spreading as a function of time.

As for the review on disorder in quantum systems, beginning in [24] the authors study the transport efficiency of an excitation moving from a source via a network to a drain. The model considered is a topologically disordered network with long-range interactions of dipole-dipole type. The authors show the crossover between purely quantum mechanical transport and environmentally induced diffusion, by phenomenologically modeling the system using quantum stochastic walk. In [25], the authors study quantum walks where signals can jump to a distant location, which is a generalization motivated by Levy flights in classical mechanics. In particular, they study two classes of quantum walks with disordered connections between beam-splitters. In the particular case of dynamic disorder, the model considered shows that decoherence leads Gaussian distribution modulated by residual patterns of quantum walk or by valleys. In [26], the authors investigate excitonic transport in systems consisting of rings of chromophores stacked in cylindrical arrays, as a function of the number of chromophores per ring, the spacing between rings, and the strength of decoherence and disorder. Using the symmetries of the system, the authors perform simulations to capture the dynamics of excitonic diffusion in the presence of environmentally-induced noise and disorder. In particular, the authors provide clear evidence for the presence of supertransfer in the appropriate regimes and for the destruction of supertransfer in other regimes. In [27], the authors investigate one-dimensional discrete time quantum walks with spatially or temporally random defects as a consequence of interactions with random environments. In particular the authors show that quantum walks with spatial disorder exhibit delocalization behaviors. In [28], the author study the discrete-time quantum walk model with Hamiltonian form of the evolution operator for each step. In particular, studying the walk dynamics using temporal, spatially static, and fluctuating disordered unitary evolutions, it is shown that localization only occurs with spatially static disordered operations. Anderson localization usually emerges in quantum systems when randomized parameters cause the exponential suppression of motion. In [29] this phenomenon is considered using the toric code. The authors show that magnetic field perturbations on the toric code induce quantum walks of anyons, which quickly destroy any stored information when anyons are present. In particular, they show that disorder induces exponential localization which suppresses the anyon motion. In [30], the authors study how disorder and fluctuations in a periodic lattice can influence the evolution of a transversing particle. In particular they show a fast ballistic spread for slowing changing lattice parameters, a diffusive spread in the case of dynamical disorder, and Anderson localization for lattices with static disorder. In [31] the authors study a spin (one-half)-particle on a one dimensional lattice subject to disorder induced by a random, space-dependent quantum coin. The discrete time evolution is given by a family of random unitary quantum walk operators, where the shift operation is assumed deterministic. Sufficient conditions on the probability distribution of the coins such that the system exhibits dynamic localization is derived. In [32], the author present an approach to induce localization of a Bose-Einstein condensate in a one-dimensional lattice under the influence of unitary quantum walk evolution using disordered quantum coin operation. It is shown that the discrete-time quantum walk on a two-state particle in a one-dimensional lattice can be diffused or strongly localized in position space, respectively. In addition, it is shown that these behaviors of the discrete time quantum walk can be efficiently induced without introducing decoherence into the system. In [33] the authors consider percolation lattices, as a simple example of a disordered system, in which edges or sites are randomly missing, interrupting the progress of the quantum walk. In one dimension quantum tunneling is used study the properties of the quantum walk as it spreads, whilst in two dimensions, it is shown that spreading rates vary from linear in the number of steps down to zero, as the percolation probability decreases towards the critical point. In [34] the dynamics of finite-sized disordered systems is considered, using the mapping between any master equation satisfying detailed balance and a Schrodinger equation in configuration space, the authors compute the largest eigenvalue relaxation time of the dynamics via lowest energy vanishing eigenvalue of the corresponding quantum Hamiltonian. In [35] the fate of quantum walks in a random environment is studied, with both static and dynamic disorder. It is shown that static disorder is responsible for exponentially suppressing quantum evolution with variance reaching a time-independent limit for long times, depending on the strength of static disorder and space dimensionality. For dynamic disorder, by coupling the quantum system to a random environment it is shown that decoherence occurs and quantum physics becomes classical so that a quantum walk is still propagating but only diffusively. In [36] the effect of static disorder on the coherent exciton transport by means of discrete Wigner functions is analyzed. It is shown that the Wigner function shows strong localization about the initial node. Integrating out the details of the time evolution by considering the long time average of the Wigner function, it is shown that localization is even more pronounced. In [37] the authors study the effect of random and aperiodic environments on cooperative processes in one space dimension. It is shown that at the critical point, both for the transverse-field Ising model and for the diffusion process, the two types of inhomogeneities have quite similar consequences, which is based on the same type of distribution of the low energy excitations. Finally in [38], the authors study controllability of a closed quantum system whose dynamical lie algebra is generated by adjacency matrices of graphs. The key property is a novel graph-theoretic feature consisting of a particularly *disordered* cycle structure. The main result is characterizing a large family of graphs that give a pair of Hamiltonians implementing any quantum dynamics, thereby rendering a system controllable.

This paper is organized as follows. In Section II we define the disordered quantum walk which can partly be found in [22]. In Section III we present our main result with proof, the weak limit theorem for the disordered quantum walk. Section IV is devoted to an open problem.

**II.  Definition of the Disordered Quantum Walk**

Consider the time evolution of the quantum walk governed by the following infinite random unitary matrices $\{U_n : n = 1, 2, \cdots\}$, $U_n = \begin{bmatrix} a_n & b_n \\ c_n & d_n \end{bmatrix}$, where the entries of the matrix are complex numbers, and the subscript $n$ denotes the time step. The unitary of $U_n$ gives

$|a_n|^2 + |c_n|^2 = |b_n|^2 + |d_n|^2 = 1$, $a_n \bar{c}_n + b_n \bar{d}_n = 0$, $c_n = -\Delta_n \bar{b}_n$, $d_n = \Delta_n \bar{a}_n$, where $\bar{z}$ denotes complex conjugation, and $\Delta_n = \det U_n = a_n d_n - b_n c_n$ with $|\Delta_n| = 1$. Put $w_n = (a_n \ b_n \ c_n \ d_n)$. Let $\{w_n : n = 1, 2, \cdots\}$ be independent and identically distributed on some space with

$E(|a_1|^2) = E(|b_1|^2) = \frac{1}{2}$ and $E(a_1\bar{c}_1) = 0$. By using the fact that $|a_n|^2 + |c_n|^2 = |b_n|^2 + |d_n|^2 = 1$,

$a_n\bar{c}_n + b_n\bar{d}_n = 0$, $c_n = -\Delta_n\bar{b}_n$, $d_n = \Delta_n\bar{a}_n$, we also see that $E(|c_1|^2) = E(|d_1|^2) = \frac{1}{2}$ and

$E(b_1\bar{d}_1) = 0$. The set of initial qubit states for the quantum walk is given by

$\Phi = \{\varphi = [\alpha \; \beta]^T \in C^2 : |\alpha|^2 + |\beta|^2 = 1\}$, where $T$ means the transposed operator. Moreover, we

assume that $\{w_n : n = 1, 2, \cdots\}$ and $\{\alpha, \beta\}$ are independent. The quantum walk governed by the above process is what is called the *disordered* quantum walk.

In this paper we will consider *disordered quantum walks of type II* which are quantum

walks described by the above process with the additional requirement that $E(|\alpha|^2) = \frac{1}{2}$, and

$E(\alpha\bar{\beta}) = 0$. In particular we will take $U_n = \frac{1}{\sqrt{2}}\begin{bmatrix} 1 & e^{i\theta_n} \\ e^{-i\theta_n} & -1 \end{bmatrix}$. Notice we can write $U_n = P_n + Q_n$, where $P_n = \frac{1}{\sqrt{2}}\begin{bmatrix} 1 & e^{i\theta_n} \\ 0 & 0 \end{bmatrix}$ and $Q_n = \frac{1}{\sqrt{2}}\begin{bmatrix} 1 & e^{i\theta_n} \\ e^{-i\theta_n} & -1 \end{bmatrix}$, then it is seen that the evolution of the walk is determined by $|\Psi_{t+1}\rangle = \sum_{x \in Z} |x\rangle \otimes (P_t|\psi_t(x+1)\rangle + Q_t|\psi_t(x-1)\rangle)$, where we have let $|x\rangle$ be an infinite components vector which denotes the position of the walker. Here the $xth$ component of $|x\rangle$ is 1 and the other is zero, further $|\psi_t(x)\rangle \in C^2$ is the amplitude of the walker in position $x$ at time $t$, and $C$ is the set of complex numbers. Let $\||y\rangle\|^2 = \langle y|y\rangle$, then the probability that the quantum walker $X_t$ is at position $x$ at time $t$ is defined by $P(X_t = x) = \||\psi_t(x)\rangle\|^2$. The Fourier transform $|\hat{\Psi}_t(k)\rangle$ of $|\psi_t(x)\rangle$ is defined as $|\hat{\Psi}_t(k)\rangle = \sum_{x \in Z} e^{-ix}|\psi_t(x)\rangle$. By the inverse Fourier transform we have $|\psi_t(x)\rangle = \int_{-\pi}^{\pi} e^{ikx}|\hat{\Psi}_t(k)\rangle \frac{dk}{2\pi}$. The time evolution of $|\hat{\Psi}_t(k)\rangle$ is

$|\hat{\psi}_{t+1}(k)\rangle = \hat{U}(k)|\hat{\psi}_t(k)\rangle$, where $\hat{U}(k) = M(k)U_n$, and $M(k) = \begin{pmatrix} e^{ik} & 0 \\ 0 & e^{-ik} \end{pmatrix}$. By induction on $t$,

we get $|\hat{\psi}_t(k)\rangle = \hat{U}(k)^t |\hat{\Psi}_0(k)\rangle$. In particular, the probability distribution can be written as

$$P(X_t = x) = \left\| \int_{-\pi}^{\pi} \hat{U}(k)^t |\hat{\Psi}_0(k)\rangle e^{ikx} \frac{dk}{2\pi} \right\|^2.$$

### III. Main Result

Consider the matrix $\hat{U}(k) = M(k)U_n$ where $M(k) = \begin{pmatrix} e^{ik} & 0 \\ 0 & e^{-ik} \end{pmatrix}$ and $U_n = \frac{1}{\sqrt{2}} \begin{bmatrix} 1 & e^{i\theta_n} \\ e^{-i\theta_n} & -1 \end{bmatrix}$.

The eigenvalues $\lambda_j(k)$, $j = 1,2$ of $\hat{U}(k)$ are given by $\lambda_j(k) = (-1)^j e^{(-1)^j w(k)}$, where $w(k)$ is determined by $\sin w(k) = \frac{\sin k}{\sqrt{2}}$. The normalized eigenvectors $|v_j(k)\rangle$ corresponding to the eigenvalues $\lambda_j(k)$, $j = 1,2$ are given by $|v_j(k)\rangle = N_j \begin{bmatrix} \frac{\sqrt{2}}{2} e^{i(\theta_n+k)} \lambda_j(k) + e^{i\theta_n} \\ 1 \end{bmatrix}$, where $N_j$ is an appropriate normalization constant. We should note that the Fourier transform $|\hat{\Psi}_0(k)\rangle$ can be expressed by the normalized eigenvectors as $|\hat{\Psi}_0(k)\rangle = \sum_{j=1}^{2} \langle v_j(k) | \hat{\Psi}_0(k) \rangle |v_j(k)\rangle$ which implies that $|\hat{\Psi}_t(k)\rangle = \hat{U}(k)^t |\hat{\Psi}_0(k)\rangle = \sum_{j=1}^{2} \lambda_j(k)^t \langle v_j(k) | \hat{\Psi}_0(k) \rangle |v_j(k)\rangle$. Using the method of Grimmett

et.al [39], we see that the $r-$th moment of $X_t$ is given by

$$E((X_t)^r) = \sum_{x \in Z} x^r P(X_t = x) = \int_{-\pi}^{\pi} \frac{dk}{2\pi} \langle \hat{\Psi}_t(k) | (D^r | \hat{\Psi}_t(k) \rangle) = \int_{-\pi}^{\pi} \sum_{j=1}^{2} (t)_r \lambda_j^{-r}(k)(D\lambda_j(k))^r |\langle v_j(k) | \hat{\Psi}_0(k) \rangle|^2 + O(t^{r-1})$$

where $D = i\left(\dfrac{d}{dk}\right)$ and $(t)_r = t(t-1)\times\cdots\times(t-r+1)$. Take the initial state as

$$|\psi_0(x)\rangle = \begin{cases} \left[\alpha e^{\frac{i\theta_0}{2}} \quad \beta e^{\frac{-i\theta_0}{2}}\right]^T, & \text{if } x = 0 \\ \left[0 \quad 0\right]^T, & \text{if } x \neq 0 \end{cases}$$. Let $h_j(k) = \dfrac{D\lambda_j(k)}{\lambda_j(k)}$, then

$$E\left(\left(\dfrac{X_t}{t}\right)^r\right) \to \int_\Omega \dfrac{dk}{2\pi}\sum_{j=1}^{2} h_j(k)^r \left|\langle v_j(k)|\hat{\Psi}_0(k)\rangle\right|^2$$ as $t \to \infty$. Substituting $h_j(k) = x$, we have

$$\lim_{t\to\infty} E\left(\left(\dfrac{X_t}{t}\right)^r\right) = \int_{-|\gamma_\varepsilon|}^{|\gamma_\varepsilon|} x^r f(x) dx$$, where $f(x) = f_K(x;\gamma_\varepsilon)(c_0 x)$ and $|\gamma_\varepsilon| = \dfrac{1}{\sqrt{2}}$. Since $f(x)$ is a

density function, we require $\int_{-\infty}^{\infty} f(x)dx = 1$ which implies that $c_0 = 1 - \left(|\alpha|^2 - |\beta|^2 + 2e^{i\theta_0}\operatorname{Re}(\alpha\bar{\beta})\right)$.

In particular we have the following

**Theorem :** $\dfrac{X_t}{t} \Rightarrow Z$, where $\Rightarrow$ means weak convergence and $Z$ has the density function

$$f(x) = (c_0 x) f_K\left(x; \dfrac{1}{\sqrt{2}}\right)$$, where $f_K(x;a) = \dfrac{\sqrt{1-|a|^2}}{\pi(1-x^2)\sqrt{|a|^2-x^2}} I_{(-|a|,|a|)}(x)$,

$$I_A(x) = \begin{cases} 1, & \text{if } x \in A \\ 0, & \text{if } x \notin A \end{cases}$$, and $c_0 = 1 - \left(|\alpha|^2 - |\beta|^2 + 2e^{i\theta_0}\operatorname{Re}(\alpha\bar{\beta})\right)$.

**IV. Open Problem**

Recall that the degeneracy of the eigenvalues is a necessary condition for localization. However, the unitary matrix governing the disordered quantum walk in this paper has none of its eigenvalues independent of $k$ in the Fourier space. Therefore it is an open problem to rigorously show non-existence of localization.